\documentstyle[a4]{article}
\newcommand{\be}{\begin{equation}}
\newcommand{\ee}{\end{equation}}
\newcommand{\bea}{\begin{eqnarray}}
\newcommand{\eea}{\end{eqnarray}}
\newcommand{\ba}{\begin{eqnarray}}
\newcommand{\ea}{\end{eqnarray}}

\def\a{\alpha}
\def\b{\beta}
\def\r{\rho}

\def\R{I\!\!R}

\def\t{\tau}
\def\f{\phi}

\def\ep{\epsilon}

\begin{document}

\begin{center}

{\large Einstein-Cartan formulation of Chern-Simons Lorentz-violating Gravity}

\vspace*{1.2cm} {\large Marcelo Botta Cantcheff $^{\ddag}$
\footnote{e-mail: botta@cbpf.br, botta@fisica.unlp.edu.ar}}

\vspace{3mm}

$^{\ddag}$ Instituto de Fisica La Plata, CONICET, UNLP \\
CC 67, Calles 49 y 115, 1900 La Plata, Buenos Aires, Argentina

%(Dated: December 21, 2007)
%Final version published in Phys. Rev. D
\end{center}
\begin{abstract}
\noindent

We consider a modification of the standard Einstein theory in four
dimensions, alternative to R. Jackiw
and S.-Y. Pi, Phys. Rev. D \textbf{68}, 104012 (2003), since it is
 based on the first-order (Einstein-Cartan) approach to General Relativity, whose gauge structure is manifest.
This is done by introducing an additional
topological term in the action which becomes a Lorentz-violating term by virtue of the dependence of the
coupling on the space-time point. We obtain a condition on the solutions of the Einstein equations, such
that they persist in the deformed theory, and show that the solutions remarkably correspond to the classical
solutions of a collection of independent $2+1$-d  (topological) Chern-Simons gravities. Finally,
we study the relation with the standard second-order approach and argue that they both coincide to leading
order in the modulus of the Lorentz-violating vector field.

\end{abstract}

\section{Introduction.}

A few years ago, a modification of Maxwell's electromagnetism
in four dimensions was proposed which considers
a kind of Chern-Simons (CS) term in the action
 $\int dx^4\, V_{\alpha}\,
\ep^{\alpha\beta \mu\nu}A_{\beta} F_{\mu \nu}$
where Lorentz symmetry is explicitly
broken by an external vector $V^\mu$
\cite{jac}. There is growing
literature on the study of this proposal and its consequences\cite{klink,consecuencias,todos}.

In a recent work \cite{yo}, we emphasized that broken Lorentz symmetry
(abbreviated as BLS) could be obtained from physically realistic background configurations
in nonlinear relativistically invariant electrodynamics. It was also pointed out that
standard Chern-Simons terms (in $2+1$-dimensions \cite{yo}) are {\it automatically
present} in a BLS action when we search for planar features (thus turning dimensional
 reduction unnecessary). In fact, the BLS action is actually a CS theory in
$(2+1)$-dimensions embedded in $(3+1)$-dimensions, and by itself, it
does not encode any information on the field-dependence in the
direction of the external (for instance, spacelike) vector $V$: if
$z$ is its affine parameter, i.e. $V=\frac{\partial}{\partial z }$,
then we get a foliation of the space-time in $(2+1)$ hypersurfaces
$\Sigma_z$ parameterized by $z$ (and $V$ is orthogonal to each
hypersurface\footnote{Notice that if the space-time (or the
space-time region considered in the integration) is simply
connected, the condition of existence of this $z$ coordinate is {\it
equivalent} to gauge invariance of the action, namely $dV=0$.}).
Therefore, the BLS action may be written as \be S_{BLS} = \int_0^L
dz \, S_{CS}[A(z) , \Sigma_z], \ee where \be S_{CS}[A(z) ,
\Sigma_z]=\int_{\Sigma_z}{\cal L}_{CS} = \int_{\Sigma_z} A(z) \wedge
d A(z) \, ,\ee is the Chern-Simons action for the 1-form gauge field
$A(z)$ on a three-dimensional manifold $\Sigma_z$. Thus, the
dependence of this field on the parameter $z$ is not determined by
this theory. It only has to satisfy usual convergence conditions.
For example, if the interval $(0,L)$ extends to $(-\infty,
+\infty)$, $A(z)$ has to be an square-integrable function ($A \in
L^2(\R)$). In this sense, we can interpret the BLS action simply as
a sum of Chern-Simons theories on manifolds $\Sigma_z$. Remarkably
notice that this describes an eventual situation of confinement of
the electromagnetic field (photon) into a $(2+1)$-manifold, which
does not result from a constraint of the charged matter into a
planar sample. The present approach actually constitutes an attempt
of naturally extending to gravity some of these ideas.

On the other hand, a Chern-Simons modification of gravity in four dimensions via a BLS term was recently introduced
 by Jackiw and Pi \cite{JPI} in a similar way as that for electrodynamics. However, this approach is based on the second
  order formulation of general relativity, where the most relevant aspects of the Maxwell theory, related to the gauge
   structure, are hidden.
This is actually the main motivation to construct an alternative formulation where the gauge structure is emphasized.
In this work, we consider a BLS/CS deformation of standard gravity but alternatively based on first-order Cartan's
 formalism (see appendix), which treats the Riemann tensor as an standard gauge curvature for the spin-connection
  which may be viewed as a gauge variable of $SO(1,3)$. Thus, such an approach is closer in spirit to the Chern-Simons
   deformation of electrodynamics \cite{jac}.

Another very important subject naturally appears in this
context: to determine the space of solutions of the deformed
theory and, in particular, under what conditions it
contains solutions of standard Einstein gravity. The question
of the persistence of the GR-solutions in the
second-order approach to CS modified gravity was analyzed
    from the beginning \cite{JPI} up to recently \cite{sols1}, \cite{sols2}.
 The role played by the
     Pontryagyn constraint (a vanishing Pontryagyn gravitational index)
     in this problem was first emphasized in Ref. \cite{sols2}, where it was observed that the satisfaction of this
      constraint is a necessary but not sufficient
     condition for these solutions.
  This issue is also analyzed in this paper
and it is shown that the problem presents some different aspects in
the present Einstein-Cartan (EC) formulation. In particular the
constraint found for
 persistent GR solutions is different here but the Pontryagyn
 constraint is also a necessary condition as in the standard
 (second-order) context \cite{sols2}. It may be argued furthermore that
 this problem (in EC) reduces to solve a collection of pure (source-free)
Chern-Simons theories.

This article is organized as follows. In Section 2 we describe and analyze the BLS/CS deformation of
 the Einstein-Cartan gravity, as well as some interesting features of the model. In Section 3,
we study the persistence of the standard GR solutions and observe
the relation of this problem with pure Chern-Simons theories in
$2+1$-dimensions. In Section 4, we discuss the relation between the
Einstein-Cartan formulation with the standard second-order approach
\cite{JPI}. Final remarks are given in Section 5.

\section{Chern-Simons modified gravity.}

The model we are going to consider here assumes a non-linear (but {\it
relativistic}) dynamics which induces a modification of
this kind (BLS) on the standard Einstein theory \cite{JPI}\cite{BLSgrav}. In this sense,
it may furthermore be argued that BLS/CS does not need to be introduced {\it by hand},
but it can naturally appear in some realistic physical situations;
for example, according to the philosophy adopted for electrodynamics \cite{yo}, in the presence of
 background gravitational fields and/or when nonuniform distributions of matter are considered.

We use both the abstract index notation\footnote{Abstract index notation is a mathematical notation for tensors
 and spinors,
 which uses indices to indicate their type. Thus the index isn't related to any basis or coordinate system.}
  (see appendix for more details), and forms notation (by omitting abstract subindices) whenever it is convenient.
   So, greek indices $\mu, \nu, ...$ \footnote{Which are raised and lowered with the Minkowski metric
    $\eta_{\mu \nu}$.} denote the element of a tetrad (vierbein) basis $(e_a)^\mu$, and consequently
     components of any tensor in this basis.

Let us propose a Chern-Simons modification of general relativity
(GR) in the first-order formalism (see appendix):

\be\label{action}
S[e,w,\f]= \frac{1}{2\kappa^2}\int_M  dx^4 \left( e^\mu \wedge e^\nu \wedge {}^*R_{\mu \nu} \; -  \;  \tau \; R^\mu_{\;\; \nu} \wedge R^\nu_{\;\; \mu} \right)+ S_{matter} [\phi]
\ee
where the two-form $R^\mu_{\;\; \nu}= dw^\mu_{\;\; \nu} + w^\mu_{\;\; \a} \wedge w^\a_{\;\; \nu}$ is defined as
 the $SO(1,3)$ field strength for the gauge field $ w^\mu_{a\;\; \nu} $.
The scalar $\tau$ is, in principle, a pointwise function of the geometry observables, as the curvature tensor,
 and of some "extra" (matter) field, denoted by $\f$. So, the embedding variable is considered itself as a
  dynamical variable rather than a fixed external quantity.

Notice then that Lorentz symmetry is preserved in a fundamental sense. If one assumes that a more fundamental unified
 theory of matter and gravity is non-linear, a saddle point expansion about background solutions typically shall give
  origin to a BLS term (and even spontaneous BLS terms) with a fixed $\tau$ of this form \cite{POS}. This may be easily argued for
   sufficiently generic non linear (toy) theories, in similar ways as that for Electrodynamics (see Ref. \cite{yo}).

The first term corresponds to the usual General Relativity (GR)
action in the Einstein-Cartan representation, the
 second one is the Chern-Simons modification, where we have assumed that the coefficient $\tau$ may depend on the
  curvature components and/or other (matter) fields. In such a sense, this term should be viewed as an interaction term.
This may be expressed as \be\label{CS-4d} S_{BLS/CS}=  2\int_M  dx^4( d\tau \wedge {\cal L}_{CS} )\ee
where  \be K^a \equiv ({}^* {\cal L}_{CS})^a \equiv \ep^{abcd} \left( w^\mu_{b\;\; \nu} R^\nu_{cd\;\; \mu} -
  \frac{1}{3}w^\mu_{b\,\,\nu} w^\nu_{c\,\,\a} w^\a_{d\,\,\mu} \right)\ee is the Chern-Simons current density
   whose divergence is the topological number called the gravitational Pontryagyn density,
 $P\equiv  {}^*R R \equiv (\ep^{abcd} R^\nu_{ab\;\; \mu} R^\mu_{cd\;\; \nu})$.

\vspace{0.7cm}

For simplicity, let us restrict ourselves to the case when $\tau$
does not depend on the geometric variables $e_\mu^a, w^\mu_{a\;
\nu} $.
Notice, remarkably, that the matter fields are coupled to
the geometry through the topological term. The third term
of (\ref{action}) encodes the dynamics of the field $\f$ but we do
not give here any explicit Lagrangian \cite{yo} \cite{Vdinam}. 
 However, we can notice that, in general, the gravitational Pontryagyn density
constitutes a source (which is a topological charge) for the
equation of motion of $\f$, i.e.,
 \be
\frac{1}{\tau'(\f)}\left( \nabla_a \left(\frac{\delta {\cal L}_{matter}}{\delta \nabla_a \f}\right) -
 \frac{\delta {\cal L}_{matter}}{\delta \f}\right)= P
 \ee
where we have assumed that $\tau$ is only a pointwise function of $\f$ but not of its derivatives.

In particular, if we consider the simplest case, where $\tau
\equiv \f$, and $S_{matter}[\f]$ is a Klein-Gordon field on a
curved space-time: \be\label{scalar} (\nabla^a \nabla_a - m^2) \f
= P \; . \ee

\vspace{0.7cm}

Let us notice that if $S_{matter} \equiv 0$, by varying the action
with respect to $\f$ one obtains an additional
 equation of motion which constrains  the space-time geometry, the Pontryagyn constraint:
\be\label{Pont-constr}
P=0\,\,.
\ee
Because this action is dipheomorphism invariant, the Einstein tensor $G^{a}_\mu$, defined in the EC approach
 as the variation of the action with respect to tetrad, is divergence free in this case. Simultaneously, so
  as in the second-order formulation \cite{JPI}, one may verify here that
$\nabla_a G^{a}_\mu \propto P \,e^a_\mu\partial_a \t$, when the CS
contribution to the covariant divergence (through the equation of motion for the spin
connection found below) is taken into account.
 Therefore, the Pontryagyn constraint implies that
this divergence vanishes \cite{sols2}. In contrast, if one adopts
a more \emph{genuine} BLS point of view, where $\t$ is assumed to
be an external
 arbitrary function of the space-time point (a background field), in principle this constraint could not be
  satisfied, and consequently, the conservation of energy-momentum of the system would be also violated.
   However there is no conceptual problem with this fact, which is
consistent
     with translation/boost symmetry violation caused by the presence of the BLS-external
     field. So the Pontryagyn constraint must be imposed if one requires that this
      symmetry be respected by the theory.

\vspace{0.7cm}

Let us now derive the equations of motion for the geometry.
Varying the action with respect to $e^\mu_{a}$, we have:
\be\label{einstein} e_\mu^a R_{ab}^{\mu \nu} = \kappa^{2}
{T'}_b^\nu = \kappa^2  ~ e^{\nu~\,a}~ {T'}_{ab} \ee where one has
defined ${T'}_{ab} := T_{ab} + g_{ab} (T_{cd}g^{cd})/2$, $T_{ab}$
being the energy momentum tensor, and the constant $\kappa$ is
related to the gravitation constant $G$ by $\kappa^2=8\pi G$, defining the torsion as \be\label{torsiondef} \Theta^\mu =D \wedge
\,e^{\mu}= d \wedge e^\mu + w^\mu_{\;\; \nu} \wedge e^\nu \ee
which vanishes in the standard formulation, constituting the
second Einstein-Cartan equation. Here, varying the modified action
with respect to $w^\mu_{a\;\; \nu} $, we obtain the equation
\be\label{torsioneq-prev} D\wedge {}^*\left(e^{\mu} \wedge
e^{\nu}\right)= (2 \kappa^2) \;2 d\tau \wedge R ^{\mu \nu}.
 \ee
 The totally antysimmetric tensor defined in the tangent space,
 may be expressed as $\ep^{\mu \nu \a \b}={}^*\left(e^\mu \wedge e^\nu \wedge e^\a \wedge e^\b \right)$. Using this and multiplying
  both hand sides by $e_\a \; e_\b$, one may finally express the equation of motion
   (\ref{torsioneq-prev}) in terms of the torsion tensor as follows:
\be\label{torsioneq} \ep^{\mu \nu}_{ \;\;\;\;\; \a \b}\;\; e^{\a}
\wedge \Theta^{\b}= 2\kappa^2\; d\tau \wedge R ^{\mu \nu} .
 \ee
This determines the effect of the Chern-Simons deformation on the space-time geometry, through an effective
 contribution to the torsion which depends on the external field. So equations (\ref{einstein}), (\ref{torsioneq})
  describe the deformed geometry in the Einstein Cartan formulation. The same equations of motion
  might have been obtained directly
   by writing the Einstein-Hilbert term of the action as\be
   S_{EH}[e,w]=\frac{1}{2\kappa^2}\int_M  dx^4 \;\ep_{\mu \nu \a\b}\; e^\mu \wedge e^\nu \wedge R^{\a\b} \; ,\ee
   which is convenient for some purposes. The corresponding vacuum Einstein equation is obtained by varying this action
    with respect to $e^\mu$, and it may be
     expressed as \be\label{einstein-forms}\ep_{\mu \nu \a\b} \;\;e^\nu \wedge R^{\a\b} =0 \; .\ee

\vspace{0.7cm}

We would like to end this part by pointing out some interesting features of this deformed theory. The gradient of the
 external field  $\tau$ dictates the coupling of the geometric degrees of freedom with
 the $SO(1,3)$ Chern-Simons three-form Lagrangian \be {\cal L}_{CS} =  w_{\mu \nu}\wedge R^{\mu \nu} -
  \frac{1}{3}w^\mu_{\,\,\nu}\wedge w^\nu_{\,\,\a}\wedge w^\a_{\,\,\mu} \equiv w \wedge R -
   \frac{1}{3}w \wedge w \wedge w.\ee
In fact, this may be expressed as $\nabla_a\tau \equiv g\, V_a $ ( $\Rightarrow \,g \equiv \left|d\tau \right|\geq 0$)
 where $V$ is a unit vector in the gradient direction. In the limit $g\to 0$ the standard torsion-free Einstein theory
  is recovered and, on the other hand, when $g \to \infty $, the CS term governs the action. In fact, notice that if
   $g$ is considered constant and we rescale the spin-connection and define the new gauge variable
    $A^{\mu \nu}\, \equiv\, \sqrt{g}\, \,w^{\mu \nu} $ and the field strength
$F^{\mu \nu} \equiv dA^{\mu \nu} + g^{-1/2} \, A^{\mu \a}\wedge  A^{\b \nu} \eta_{\a \b}$,
the action (\ref{action}) may be written as (from now on, we set $2\kappa^2 =1$):
\be\label{actionA}
S_{Grav}[e,A,\f]= \int_M  dx^4 \left(g^{-1/2}\, e^\mu \wedge e^\nu \wedge {}^* F_{\mu \nu} \; + \frac12
  \; V \wedge \left(A \wedge F - \frac23 A \wedge A \wedge A \right)\right) ,
\ee
where we have used the equivalence of the second term of (\ref{action}) with the Chern-Simons form.
Thus, we can see in this expression that, in this case, the first term is a first-order perturbation in
 $(\sqrt{g})^{-1}$ while the second one, the Chern-Simons action, may be seen as the free kinetic term
  \footnote{In this sense, we would
like to mention the possibility of recovering a theory with local degrees of freedom
 from a topological theory through a perturbative method \cite{POS}, in the sense of Ref. \cite{freidel} and
also \cite{rov}.}.
  On each level (hyper)surface of the
    field $\tau(x)$, we have a Chern-Simons action for the connection $A^{\mu  \nu}$ in the group $SO(1,3)$,
     which contains the Lorentz-Poincar\'e group $ISO(1,2)$ if the dreibein
      $E^{\hat\mu } \,\,\, ( {\hat\mu },{\hat\nu }=0,1,2 ) $ , the gauge field associated with translations
       on those hypersurfaces, is identified with $A^{{\hat\mu } , 3}$ and the spin-connection
        $A^{{\hat\mu }{\hat \nu}}$ is the gauge field associated with $SO(1,2)$. This theory precisely describes
         3d-gravity, which
        is exactly soluble (there are no local
   degrees of freedom) and
  its quantization is well understood \cite{witten}.
By a similar argument as that for electrodynamics (shown in the introduction), we may observe that in the large
 $g$ limit, the theory becomes
 \textit{a collection} of decoupled Chern-Simons  gravities on 2+1-dimensional manifolds which foliate the space-time.
 The function $\t$ parameterizes these hypersurfaces and $g=\left|\tau'\right|$ encodes their density/number.
  So when this number is large the theory approaches to the CS description \footnote{This might be interpreted
   as a sort of macroscopic limit where the microscopic component are $2+1$ dimensional manifolds equipped with
    CS theories. The reader may find some close perspectives in Ref. \cite{micro-st}}.
 Apart from this, in the next section we are going to show that precisely these planar Chern-Simons theories describe
  the Einstein persistent solutions of the theory.

All these features naturally suggest an important question: Could this strong/weak behavior be interpreted as
 duality in some proper sense?
Clearly, the answer could have some relation with the paradigmatic holographic principle (t'Hooft 1993 and
 Susskind 1995) \cite{holography}
 and it shall be carefully analyzed elsewhere \cite{prep}.

\section{BLS/CS Deformation and Persistence of Solutions.}

Let us study some remarkable aspects of the problem of the persistence of the solutions in the
 Einstein-Cartan formulation of BLS/CS gravity.
Consider the decomposition of the curvature \be\label{decF} R
^{\mu \nu} = r^{\mu \nu} \wedge V + \hat{F}^{\mu \nu} .\ee Let us
assume that $\tau$ parameterizes a foliation of the space-time $\{
\Sigma_\t\}_\t$, thus we may define the projector $h\equiv g - V
\otimes V$ ($h\equiv g + V \otimes V$, if $V$ is timelike) on each
hyper-surface of the foliation,
%\footnote{One always may consider at
%least regions of the space-time where this is well defined.},
   then $\hat{F}^{\mu\nu}_{ab} \equiv h^c_a\, R^{\mu \nu}_{cd}\, h^d_b \equiv h\, R\, h$. Therefore, Eq.
    (\ref{torsioneq})
may be expressed as: \be \ep^{\mu \nu}_{ \;\;\;\;\; \a \b}\;\;
e^{\a} \wedge \Theta^{\b}= d\tau \wedge \hat{F}^{\mu \nu}. \ee If
we also consider the spin-connection one form decomposition
$w^{\mu\nu} \equiv \a^{\mu\nu} V
 + \hat{w}^{\mu\nu}$, where $\hat{w}_a^{\mu\nu}\equiv h^c_a \,w_b^{\mu\nu}\equiv h \, w^{\mu\nu}$,
  and use $dV=dd\t= 0$, one may verify that $\hat{F}^{\mu \nu}$ is the curvature corresponding to
   the connection $\hat{w}^{\mu \nu}$.

 Notice that the theory is torsion-free if and only if the connection $\hat{w}^{\mu \nu}$, defined
  on the 2+1-embedded surfaces and valued on the de Sittergroup in 2+1-dimensions, $SO(1,3)$, is such that
   the associated (three-dimensional) curvature vanishes, which reveals an interesting structure
    related to the homotopic classes.
 So, the condition for the persistence of EC solutions reads
\be\label{persistence}
  \hat{F}^{\mu \nu} =0,
 \ee
 where $\hat{F}^{\mu \nu}$ is the curvature of the gauge variable corresponding to the de Sitter group
  of the $\Sigma_\t$-submanifolds.
Therefore, such connections
 on appropriate foliations, represent torsion-free geometries, and furthermore (remarkably), the solutions
 coincide with those of standard Einstein theory.
Then, for each solution of (\ref{persistence}), a pure gauge, one has a persistent solution.
 They may be expressed as
\be
\hat{w}^{\mu}_{\;\nu} \,= \, G^{\mu \a} \,d \,G^{-1}_{\a\nu}, \,\,\,\, , \,\,\,\, G\in SO(1,3).
\ee
Finally, one may use this form in the EC equations, and in this way, \emph{to construct} all
 the GR (torsion-free) preserving solutions. Therefore, we may
 remarkably notice the existence of a
correspondence between the classical solutions of pure
source-free Chern-Simons theories (defined on a collection of
2+1 dimension manifolds which foliate the space-time) and the solutions
of standard Einstein gravity, provided that they are solutions of
full theory (\ref{action}).

Notice that the present preserving condition is stronger than the Pontryagyn constraint
 $P=0$.  In fact,
by using (\ref{decF}), we get \be\label{pontr-cond}
 P =  {}^{*} (R_{\mu \nu}\wedge R^{\mu \nu} )= {}^{*} ((r_{\mu \nu} \wedge V + \hat{F}_{\mu \nu})\wedge
  (r^{\mu \nu}\wedge V + \hat{F}^{\mu \nu} ))= {}^{*} (2 r_{\mu \nu} \wedge V \wedge \hat{F}^{\mu \nu}
   + \hat{F}_{\mu \nu}\wedge \hat{F}^{\mu \nu} ),
 \ee
 which vanishes for the solutions of (\ref{persistence}). In agreement with this, in Reference  \cite{sols2} it was
  already found that the Pontryagyn constraint
 is a necessary but not sufficient condition for
 persistence of solutions of the theory in the second-order formulation.
This is important to check out
 consistency with the vanishing of the divergence of the energy-momentum tensor discussed
  in the previous Section, even when the field $\t$ is considered external. In fact, for
   (vacuum) persistent solutions, the Pontryagyn constraint is satisfied, and therefore,
    the Einstein tensor is divergence free as expected.

    The question of the persistence constitutes an appropriate ambient to
    discuss the relation of this approach with the standard second
    order formulation \cite{JPI}, since in both, it reflects the
    contribution to the equation of motion of the CS deformation.
    In fact in the second-order formulation, the
    persistence is ruled out by a vanishing Cotton tensor, which
    in standard notation is
    expressed as
   \be\label{cotton}
   C={\bar \nabla}_c \t \; \ep^{cde(a}\nabla_e R^{b)}_{\;d} +
   ({\bar \nabla}_{(c}{\bar \nabla}_{d)}\t) \;{}^*R^{d(ab)c}
   \ee
   where ${}^*R_{\;\,b}^{a\;\;ef}= \frac12
   \ep^{efcd}R_{\;\,bcd}^{a}$, and this is related to the curvature
   tensor in the tetrad notation as $R_{\;\,bcd}^{a} \; e_a^{\mu} e^{b\,\nu}= R_{cd}^{\mu \nu}$
   for a given external field $\t$. This curvature is associated to the canonical covariant derivative,
    ${\bar \nabla}_a$, which is torsion-free and compatible with the metric $g_{ab}$.
    The precise relation between these conditions of persistence will become clear in the next section.

   Finally, we would like to remark that in nearly flat regions of the space-time (e.g. the spacial infinity
   of asymptotically flat solutions) the
GR solutions are preserved independently of the magnitude of $g=\left| d\t \right|$. In particular, for all
 asymptotically flat space-time of the undeformed GR theory, the right-hand side of Eq. (\ref{torsioneq}) vanishes,
  and BLS is undetectable near of the spacial infinity.

\section{Einstein-Cartan Approach vs the Standard Formulation}

Solving Eq. (\ref{torsioneq}) [using (\ref{torsiondef})] for
the spin coefficients $w^\mu_{a\;\; \nu}$ in terms of $ e^\nu_a$ and $\partial_a \tau$
  and replacing the solution into (\ref{einstein}), we recover the modified Einstein
  equation for the tetrad $ e^\nu_a$ (or, equivalently, for the metric $g_{ab}$)
which may be seen as the equation of motion of a formulation of
the theory whose only variable is the metric, however, this is
\emph{a priori} inequivalent to the standard second-order
Jackiw-Pi approach \cite{JPI}. Because of the presence of the torsion
in this description one may trivially argue that the geometries
described by the solutions of both formulations are very
different. However here, we are going to discuss this question more carefully.

\vspace{0.5cm}

Let first us show that in fact one can solve Eq. (\ref{torsioneq}) and find out a
solution for $w^\mu_{a\;\; \nu}$ in terms of $ e^\nu_a$ and
$\partial_a \tau$ even in modified gravity. We may do that by
constructing a sort of perturbation scheme in the deformation
parameter $g$, where each order in the expansion may be
iteratively solved in terms of the lower ones. It shall be
emphasized, however, that this procedure, developed here to study some properties
of this formulation and its relation with the second-order formalism, should not be seen as a
method to solve the equations of motion since it generates an
equation for the tetrad whose order, in principle, grows as the power of $g$, which
 would require a consistent truncation to be solved. Because of this, it is
convenient to solve the Eqs. (\ref{einstein})
(\ref{torsioneq}) as a first-order system of coupled equations.

\vspace{0.5cm}

Consider the solution of Eq (\ref{torsioneq}) to be $w^{\mu \nu}=
W^{\mu \nu} + K^{\mu \nu}$ where $W^{\mu \nu}$ is the undeformed
torsion-free (Christoffel) spin connection and $K^{\mu \nu}$ is
the contortion one-form, then \be\label{ultima} \theta^{\mu}=
K^{\mu \nu}\wedge e_{\nu} .\ee Substituting this into Eq.
(\ref{torsioneq}) we obtain
\be \ep^{\mu \nu}_{ \;\;\;\;\; \a \b}\;\; e^{\a} \wedge K^{\b
\r}\wedge e_{\r} = g V \wedge \left( R^{\mu \nu}[W]+
W^{\mu}_{\;\;\a} \wedge K^{\a \nu} + K^{\mu}_{\;\;\a} \wedge W^{\a
\nu} + R^{\mu \nu}[K]\right) ,\ee where \be R^{\mu \nu}[W]=
dW^{\mu \nu}+ W^{\mu}_{\;\;\a} \wedge W^{\a \nu} ~,~~~~~~~R^{\mu
\nu}[K]= dK^{\mu \nu}+ K^{\mu}_{\;\;\a} \wedge K^{\a \nu}\,\,.\ee

Let us consider now a solution $K$ being an analytic function of $g$,
which here, is assumed to be constant for
simplicity.
%\footnote{However, it will be finally observed that the
%existence and the method to find the solution not depend on it.}.
By consistency with the definition we clearly see that $K(g=0)=0$. The
zeroth order equation is $\Theta=0$, which may be solved in terms
of the frame and its partial derivatives\be\label{solW} W^{\mu
\nu}= f^{\mu \nu}(e, \partial_a e)\ee Then, let us consider
Taylor's expansion in powers of $g$: \be\label{e27} K^{\mu \nu}\equiv
\sum_{n=1}^\infty \; g^n \; k_n^{\mu \nu}~,~~~~~~~~~\Theta^{\mu
}\equiv \sum_{n=1}^\infty \; g^n \; \theta_n^{\mu }
~,~~~~~~~~~\theta_n^{\mu } \equiv k_n^{\mu \nu}\wedge
e_{\nu}~.~~~~~~~~~~~ \ee Substituting this into Eq.
(\ref{torsioneq}), we get to first-order \be\label{k1} \ep^{\mu \nu}_{
\;\;\;\;\; \a \b}\;\; e^{\a} \wedge k_1^{\b \r}\wedge e_{\r} =  V
\wedge R^{\mu \nu}[W] + o(g) \ee which may be easily solved for
$k_1$ (or $\theta_1$)
 in terms of $W^{\mu \nu}, dW^{\mu \nu}$, and $e^\mu$, \footnote{The solution reads  $
\theta_{1~ab}^{\mu}= -\frac34 \,  \ep^{\mu \nu}_{ \;\;\;\;\; \a
\b}(V \wedge R^{\a \b}[W])_{abc} \, e^c_\nu + o(g)$. } which
furthermore by virtue of (\ref{solW}) may be expressed in terms
of $e^{\mu }$.
%\footnote{It actually shall read as a function of
%$e^{\mu }$ and its (first or upper order) derivatives.}.
Finally, one may use the same procedure iteratively; order by order, the
right hand side of the resulting equation will depend on the lower
ones, namely, \be \ep^{\mu \nu}_{ \;\;\;\;\; \a \b}\;\; e^{\a}
\wedge k_{n+1}^{\b \r}\wedge e_{\r} = V \wedge \left( d k_n^{\mu
\nu} + W^{\mu}_{\;\;\a} \wedge k_n^{\a \nu} + k^{~\;\mu}_{n\;\;\a}
\wedge W^{\a \nu} + \sum_{m=1}^{n-1}k^{~\;\mu}_{m\;\;\a} \wedge
k^{\a \nu}_{n-m}\right) .\ee Therefore, one may conclude that
$k_{n}\, ,\forall n \geq 1$ by induction, and consequently the
full connection $w^{\mu \nu}$, may be expressed in terms of
$e^{\mu}$ as claimed above. Notice that only at the trivial order
($g \to 0$), the corresponding deformed Einstein equation results
to be a second-order equation in partial derivatives of the
variable $e^{\mu}_a$ (or $g_{ab}$). In principle, higher powers in $g$
generically would contribute with higher order derivatives to this
equation; however, it is possible that derivatives of the tetrad fields of orders
 higher than 3 in the deformed Einstein equation may be eliminated by using the
  Bianchi or other identities.
  %A general calculation in this sense is quite complicated technically
  % and we will not deal with it in this article, however we are able to clarify the relation
  %  of the present formulation with the third order (in the tetrad field) equations of motion
  %   of the standard formulation \cite{JPI}.
A general calculation in this sense is a bit complicated technically
   and not very illuminating for our purposes here. We are able to clarify, however, the relation
    of the present formulation with the third order (in the tetrad field) equations of motion
     of the standard formulation \cite{JPI}.

Plugging the solution $w^{\mu \nu}=
W^{\mu \nu}(=f^{\mu \nu}(e, \partial_a e)) + g k_1^{\mu \nu} + \dots$, back into (\ref{action}), we obtain the
 CS deformed action for the tetrad field. Considering up to the first-order in $g$, we may express this as:
\be\label{actione}
S[e]= \int_M   e_\mu \wedge e_\nu \wedge {}^* \left(R^{\mu \nu}[W] + g D_W \wedge k_1^{\mu \nu} \right) \; +
  \;  \int_M  g \; V \;\wedge {\cal L}_{CS}[W] + o^2(g) \; ,
\ee
where $W$ is the (torsion-free) Christoffel connection expressed in terms of the tetrad [Eq. (\ref{solW})] and
 $D_W$ is the correspondent covariant derivative. The second term may then be integrated by parts and expressed
  as $g\int (D_W \wedge {}^*(e_\mu \wedge e_\nu))\wedge k_1^{\mu \nu}$ up to boundary terms. This finally vanishes
   due to the torsion-free condition. Therefore, by definition of the Christoffel connection (encoded in $W$),
    this action is coincident with that of Jackiw-Pi expressed in the first-order Einstein-Cartan language.
     The variation of this action with respect to the tetrad, may then be expressed as
\be\label{einstein-cotton}
R_a^{\;\,\mu} + C_a^{\;\,\mu} =0 ,
\ee
where $C_a^{\;\,\mu} $ corresponds to the variation of the last term of (\ref{action}) with respect
 to the tetrad which coincides with the Cotton tensor ($C_a^{\;\,\mu} \,e_{\mu \, b} = C_{ab}$) by definition.
The same result is obtained by plugging the first-order solution (\ref{k1}) into the (vacuum) Einstein equation
 (\ref{einstein}).

So, we may conclude that the present Einstein-Cartan  formulation of
CS modified gravity \emph{coincides} with the standard approach (Ref
\cite{JPI}) to first-order in the modulus of the breaking vector
$g(=\left| d\t \right|)$.

Notice in addition that if the constraint (\ref{persistence}) is satisfied for all order in $g$, then the
 full connection $w$ also satisfies the torsion-free condition; thus $w=W$, and $K=0$. Thus as in the procedure
  above, substituting this solution into the action (\ref{action}) gives the results:
\be\label{actione-C} S[e]= \int_M   e_\mu \wedge e_\nu \wedge {}^*
R^{\mu \nu}[W]  \; , \ee where the constraint (\ref{persistence})
was used to eliminate the last term of (\ref{actione}). The
corresponding equation of motion reduces to the vacuum Einstein
equation, $R_a^{\;\,\mu}=0$. In other words, the persistence
condition (\ref{persistence}) \emph{implies} that the Einstein
equation remains undeformed as expected. In particular to first
order in $g$, consistency with Eq. (\ref{einstein-cotton}) requires
that the Cotton tensor vanishes identically when condition
(\ref{persistence}) is satisfied.
In this way, we have used the statement on the agreement to first-order of both formulations,
 to argue that our persistence condition (\ref{persistence}) not only guarantees that the space-time is torsion-free, but 
  also that furthermore the metric satisfies the unmodified Einstein equation.

\subsection{Spherically symmetric solution and nonperturbative (in)equivalence.}

Concerning the equivalence of both formulations beyond
the first order of the $g$ expansion, we shall verify
here that the Schwarzschild solution, which is persistent in
the Jackiw-Pi formulation for a particular choice of
$d\t$, but \textit{is not} a solution of the present theory, in particular, the second-order already breaks down that persistence. This fact contradicts the nonperturbative equivalence of both formulations.

Let us consider the Schwarzschild solution given by the tetrad \cite{wald}:
\bea\label{terada-schw}
e_0 &=& f^{1/2}(r) dt \;~,\nonumber\\
e_1 &=& f^{-1/2}(r) dr \;~,\nonumber\\
e_2 &=& r\; d\theta \;~,\nonumber\\
e_3 &=& r \;sin \theta \, d\f \;~, \nonumber\\
f(r)&=& 1 - 2M/r \;,
\eea
and the particular choice $\t \equiv g_0^{-1} t$, where $g_0$ is an arbitrary constant. In Ref. \cite{JPI} it
 was shown that this is an exact solution of the theory in the standard formulation, and in Ref. \cite{sols2} it was extended to other choices of the breaking vector. In the present case this vector does not have a constant modulus;
 however, we even may define an expansion as (\ref{e27}) controlled by the parameter $g_0$. Namely, $V\equiv e^0$, $d\t = g\, V \equiv g_0 f^{-1/2} \, V $.

Equation (\ref{k1}) gives the (first-order) torsion for the Schwarzschild space-time. The right-hand side of that equation is determined by the components of the curvature orthogonal to $dt$, associated with the torsion-free connection of the Schwarzschild solution:
\bea
R^{12} &=& A(r) \,dr \wedge d\theta \;~,\nonumber\\
R^{13} &=&  A(r)\, sin \theta \,dr\wedge d\f \;~,\nonumber\\
R^{23}  &=& 2 (1-f) \,sin \theta \, d\theta \wedge d\f \;~,
\eea where $ A(r)\equiv -\frac{2M}{r^2}\, f^{-1/2}$. 
 The corresponding nontrivial contortion coefficients may be directly obtained by plugging (\ref{ultima}) into (\ref{k1}) and solving a linear algebraic system of equations. The non-trivial coefficients are
\bea\label{solucionk1}
k_{(1)}^{12} &=& \left(\frac{s_1}{2} - s_2 \right)\;e_3 \;~,\nonumber\\
k_{(1)}^{13} &=& \left(\frac{s_1}{2} + s_2 \right)\; e_2 \;~,\nonumber\\
k_{(1)}^{23}  &=& -\frac{s_1}{2} \; e_1 \;~,
\eea
where \be
s_1=\frac{2(1-f)}{r^2} = \frac{4M}{r^3}\;~\;~,\;~\;~\;~\;~
\;~s_2=\frac{f^{1/2}A}{r}=-\frac{2M}{r^3}\;~.
\ee
%Then notice that $k_{(1)}^{13}=0$.
On the other hand, the modified (vacuum) Einstein equation reads
\be
\ep_{\mu\nu \a \b}\;e^{\nu}\;\wedge\;( R^{\a \b}[W] + D_W \, K^{\a \b} + \eta_{\r \kappa} \, K^{\a \r} \wedge K^{\kappa \b})\,=\,0\;.
\ee
So, the condition for the persistence of the solution for the tetrad (\ref{terada-schw}) in the EC approach is
\be
\ep_{\mu\nu \a \b}\;e^{\nu}\;\wedge\;(D_W \, K^{\a \b} + \eta_{\r \kappa} \, K^{\a \r} \wedge K^{\kappa \b}) \,=\,0\,\,.
\ee
 The first-order of this equation is trivial since this solution is persistent in the standard approach \cite{JPI}, thus we may formulate the persistence condition for the following order as
\be\label{second-order}
\ep_{\mu\nu \a \b}\;e^{\nu}\;\wedge\;(D_W \, f^{-1} k_{(2)}^{\a \b} + f^{-1} \,\eta_{\r \kappa}\, k_{(1)}^{\a \r}\, \wedge\, k_{(1)}^{\kappa \b}\,)  + \,o(g_0) \,=\,0\;.
\ee
In fact, we are going to observe that this equation cannot be satisfied and consequently, that the Schwarzschild metric is not a solution in the EC formulation. The second-order contortion coefficients may be obtained by solving the equation
\be \ep^{\mu \nu}_{ \;\;\;\;\; \a \b}\;\; e^{\a} \wedge k_{(2)}^{\b
\r}\wedge e_{\r} =  f^{1/2} V \wedge \left( D_W \, f^{-1/2} k_{(1)}^{\mu \nu} \right) ,\ee which is similar in form to Eq. (\ref{k1}) and may be solved in the same way. Substituting $k_{(1)}$ by the solution (\ref{solucionk1}), and using that $d \, H= f^{1/2} \, H' \,e_1 \, ,\,\, \forall \,H=H(r)$, it may be easily shown that \be k_{(2)}^{1 \b}=0 \;.\label{k1nu}\ee
%the relations $D_W \, e_\mu =0$
Therefore, 
it is convenient to search for the component $\mu=2$ of the right-hand side of (\ref{second-order}) which, by virtue of (\ref{k1nu}), reduces to
\be
\ep_{2 \, 0\, 1 \,3  }\; e^{2}\;\wedge\;e^{0}\;\wedge\;(f^{-1} \, k_{(1)}^{1 \,2}\, \wedge\, k_{(1)}^{2\, 3}) \,=\,f^{-1}\;\frac{s_1}{2}( \frac12 s_1 - s_2)\,e_{0}\wedge e_1 \wedge e_2\wedge e_3= \,f^{-1}\; s_1^2 \,\,\,\,e_{0}\wedge e_1 \wedge e_2\wedge e_3 \neq 0 \;,
\ee
where we have also multiplied by $e_2$ and used that $k_{(1)}^{0 \mu}=k_{(2)}^{0 \mu}=0$ and the antisymmetry of $\ep_{\mu\nu \a \b}$. This is clearly in contradiction with the condition (\ref{second-order}). Thus, the Schwarzschild metric \textit{is not} a solution to the deformed Einstein equation in the EC approach which means  that equivalence with the standard formulation is lack. So, we may conclude this section by emphasizing that both formulations approach each other to leading order in $g$, but they are inequivalent because the contribution of the higher orders is not trivial.

\section{Final remarks}

This work consists in the natural application to gravity of some ideas
 about theories with a Chern-Simons term in four dimensions, which breaks the Lorentz symmetry through a
  formulation where the gauge structure of the theory is explicit \cite{yo}.

We found the conditions to get persistent GR solutions. They have a simple geometric interpretation and link with
 topological gauge theories. In a forthcoming paper, we will focus on the study of these and other exact solutions
  of the deformed theory.

Finally, we analyzed the relation between the present Einstein-Cartan formulation of CS-Lorentz-violating gravity
 and the standard one proposed by Jackiw and Pi \cite{JPI}, based on a Taylor expansion in powers of the
  modulus of the external breaking vector.

\section{Acknowledgements}

The author would like to thank O. A. Reula, J. A. Helayel-Neto and G. Silva for
useful comments and observations. 
This work was supported by CONICET.

\section{Appendix: The Abstract index Notation and Einstein-Cartan formalism}

In this work, we shall use the abstract index notation
\cite{wald}, namely, a tensor of type $(n,m)$ shall be
 denoted by $T^{a_1 .....a_n}_{b_1 .....b_m}$, where the Latin index stands for the numbers and
 types of variables on which the tensor acts and not as the components themselves on a certain basis. Then,
  this is an object having a basis-independent meaning.
 In contrast, Greek letters label the components, for example $T^{\mu \nu}_{\alpha}$
 denotes a basis component of
  the tensor $T^{ab}_c$. We start off with the Cartan's formalism of GR.
 We introduce \cite{wald} an orthonormal basis of smooth vector fields $(e_{\mu})^a$,
 satisfying
\be (e_{\mu})^a(e_{\nu})_a = \eta_{\mu \nu},\label{orto}
 \ee where
$\eta_{\mu \nu}=diag(-1,1,1,1)$. In general, $(e_{\mu})^a$
is referred to as {\it vielbein}.
 The metric tensor is expressed as
\be\label{defg-e} g_{ab}=(e^{\mu})_a(e^{\nu})_b \eta_{\mu \nu}  \;. \ee
From now on, component indices $\mu,\nu,..$ will be raised and lowered
using the flat metric $\eta_{\mu\nu}$ and the abstract ones, $a,b,c...$
with space-time metric $g_{ab}$.

 Now we define the {\it Ricci rotation coefficients}, or {\it
spin-connection}, \be (w_{\mu\nu})_a=(e_{\mu})^b\nabla_{a}(e_{\nu})_b \;
, \label{defw} \ee where $w_{a\mu\nu}$ is antisymmetric, and,
together with (\ref{orto}), is equivalent to the compatibility
condition\be \nabla_a g_{bc}=0 \; . \label{compat}\ee From (\ref{defw}) we have
\be \nabla_{a} \;\; e^{\mu }_{\;\;b} + w^{\mu\nu}_{\;\;\;a} \; e^{}_{\nu \, b} =
 \partial_{a} \;\; e^{\mu }_{\;\;b} + \Gamma^c_{ab} \;\; e^{\mu }_{\;\;c}+ w^{\mu\nu}_{\;\;\;a} = 0\; , \label{e-w}
\ee where $\Gamma^c_{ab}$ are the Christoffel symbols connection. It
is useful to define the part
 of the covariant derivative referred only to the internal indices correspondent to the spin-connection $w^{\mu\nu}_{\;\;\;a}$, denoted by $D_a$.

The antisymmetric part of (\ref{e-w}) (with the convention of antisymmetrization $(...)_{[ab]} = ((...)_{ab} - (...)_{ba})/2$ reads \be \nabla^{}_{[a}~e_{~~b]}^{\mu
} = - w^{\mu\nu}_{~~~[a} \; e^{\alpha}_{~~b]}\eta_{\nu \alpha} \;. \label{antis-e-w} \ee
In the standard Einstein formulation of GR, the
connection is assumed to be torsion free. This is expressed by
\be ( D\wedge e^\mu )_{ab} \equiv D^{}_{[a}~e_{~~b]}^{\mu
}=
\partial^{}_{[a}~e_{~~b]}^{\mu
} + w^{\mu\nu}_{~~[a} e^{\alpha}_{~~b]}\eta_{\nu \alpha}= 0\;\; . \label{notorsion}
\ee
The components of the Riemman's tensor in this orthonormal
basis are given as follows \be R^{~~\mu \nu}_{ab}:= 2\partial^{}_{[a}
w^{\mu \nu}_{~~b]} + 2w^{\mu \rho}_{~~[a}w^{\sigma\nu}_{~~b]}
\eta_{\rho\sigma}. \label{p1} \ee Equations (\ref{notorsion}) and
(\ref{p1}) are the {\it structure equations} of GR in Cartan's
framework.

Einstein's equation in this framework reads \be e_{\mu}^{~~a} R^{~~\mu \nu}_{ab} =
\kappa^2  ~ e^{\nu~\,a}~ {T'}_{ab},\label{eins} \ee where one has defined ${T'}_{ab} :=
T_{ab} + g_{ab} (T_{cd}g^{cd})/2$, $T_{ab}$ being the energy-momentum tensor, and the constant $\kappa$ is related to the gravitation
constant, $G$, by $\kappa^2=8\pi G$.

Equations (\ref{e-w}) and (\ref{eins}) are a system
of coupled first-order nonlinear equations for the variables $(e,w)$
which determine\footnote{Together with
the antisymmetry condition for $w_a$.} the dynamics of GR.

This yields the so-called ``Einstein-Cartan formalism"; we obtain, thereby, a first-order
Einstein-Hilbert action which can be expressed as
\be S={1 \over 2\kappa^2}\int dx^D ~e~R^{~~\mu \nu}_{ab}e_{\mu}^{~~a} e_{\nu}^{~~b}, \label{E-H}
\ee where $e=(-\det g)^{1/2}= \det (e^{\mu}_{~~a})$. If we wish to consider
a nonvanishing cosmological constant, $\Lambda$,
 $R^{~~\mu \nu}_{ab}$ must be replaced by
\be R^{~~\mu \nu}_{ab} + \Lambda e^{[\mu}_{~~a} e^{\nu]}_{~~b}.
\label{cosmol} \ee

\end{document}